\documentclass[aps,pra,epsfigure,twocolumn,showpacs]{revtex4}
\usepackage{amsmath}
\usepackage{amstext}
\usepackage{latexsym}
\usepackage{psfig}
\usepackage{graphicx}
\usepackage{amsfonts}

\newcommand{\ket}[1]{\left\vert#1\right\rangle}
\newcommand{\modul}[1]{\left\vert#1\right\vert}

\begin{document}
%\draft

%\wideabs{

  \title{Simulation of quantum random walks using interference of classical field}

  \author{H. Jeong, M. Paternostro, and M. S. Kim}

\affiliation{School of Mathematics and Physics, The Queen's University,
    Belfast BT7 1NN, United Kingdom}
   \date{\today}

%  \maketitle

\begin{abstract}
  %It has been pointed out that the distribution patterns of quantum
  %random walks, which are different from their classical counterpart,
  %are caused by quantum coherence.
  We suggest a theoretical scheme for the simulation of quantum random walks
  on a line using beam splitters, phase shifters and photodetectors. Our model enables us to simulate a quantum random walk with use of the wave nature of classical light fields.
  Furthermore, the proposed set-up allows the analysis
  of the effects of decoherence. The transition from a pure mean photon-number distribution to
  a classical one is studied varying the decoherence parameters.
\end{abstract}
\pacs{03.67.-a, 03.67.Lx, 42.25.-p, 42.25.Hz, 03.65.Yz}

%}

\maketitle
%%%%%%%%%%%%%%%%%%%%%%%%%%%%%%%INTRODUCTION%%%%%%%%%%%%%%%%%%%%%%%%%%%%%%%%%%%%
\section{Introduction}
\label{Intro}
Random walks are useful models for physicists to
study statistical behaviours of nature such as Brownian motions
of free particles~\cite{MandelWolf}. They have also been studied
for practical use such as  algorithms in computer science
\cite{Vigoda} and risk management in finance~\cite{EF}.
Quantum versions of random walks have been recently studied both
for fundamental interests and for the expectation of building new
algorithms for quantum computation~\cite{Ambainis}. There have
been several suggestions for a practical implementation of
quantum random walks, using ions in linear traps, optical
lattices and cavity-QED~\cite{Kempe2,Implementations}. Recently,
proposals for the implementation of quantum random walks with
linear optical elements have been suggested~\cite{Cinesi,bergou}
and the first search algorithm using quantum random walks has been
reported~\cite{Deotto}. Quantum random walks typically show very
different patterns from the Gaussian distributions for classical
random walks, which have some remarkable characteristics such as an
exponentially fast hitting time~\cite{Ambainis}. It has been
pointed out that these differences are due to the existence of
quantum coherence~\cite{Kempe2}.

In this paper, we suggest a theoretical scheme to simulate quantum random walks
on a line using the wave nature of classical light fields. This is related
to the fact that the idea of quantum coherence is originally borrowed from the
interference of wave mechanics shown in Young's double-slit experiment.
In our scheme, it is also possible to simulate decoherence processes modeled using
additional random phase-shifters and beam splitters with
erratic transmittivity. This analysis is relevant under different points of view. First of all it shows that, increasing the amount of decoherence that affects the system, the distribution of the random walk changes from a totally quantum one to a classical Gaussian distribution. This clarifies the role played by the interference effects in the dyanmics of a quantum walker and represents an ulterior proof of the validity of a simultation based on interferometric devices. On the other hand, studying the effects of possible sources of errors in our model, we can single out the causes of certain deviations from the ideality in the patterns resulting from performed experiments.

This paper is organized as follows. In Section~\ref{Setup}, we
briefly review coined quantum random walks on a line with their
characteristics and we suggest a scheme for the
simulation of quantum random walks. We will later show that, with this set-up,
we can simulate quantum random walks using wave nature of a field. This possibility is neither always
obvious nor mentioned in other models. It should be pointed out that a similar scheme
of all-optical implementation has been suggested by Zhao {\it et al.}~\cite{Cinesi}.
Their proposal is entirely based on the quantum coherence of a quantum superposition of
two polarization degrees of freedom. On the other hand, our scheme is able to take the wave nature of any
input field (classical or non-classical) to show the same interference pattern. This, with the explanation of the
 role of the phase shifters in our scheme, is shown in Section~\ref{general proof}.
Section~\ref{decoerenza} is devoted to the study of the
decoherence effects in our proposal. We show that decoherence on
the coin tossing operation and on the quantum walker motion can
be simulated and studied by means of our system, thus
demonstrating the role of the interference effects  in this
simulation. This investigation is useful even from a practical
point of view because it singles out and characterizes the effect
of a class of errors that could affect a performed experiment.
%%%%%%%%%%%%%%%%%%%%%%%%%%%%%%%%%%%%%%%%%%%%%%%%%%%%%SETUP%%%%%%%%%%%%%%%%%%%%%
\section{Quantum random walk with linear optical elements}
\label{Setup}

In uni-dimensional coined random walks, the walker is restricted
to move along a line with a number of discrete integer points on
it.  The walker is supposed to be a classical particle on one of
the integer points. A coin tossing determines whether the walker
moves left or right for each step. In the quantum version of
coined random walks, the classical coin is replaced by a quantum
bit whose states $\ket{L}$ and $\ket{R}$ represent the logical
values  LEFT and RIGHT. The quantum coin can be embodied by an
internal degree of freedom of the walker itself~\cite{Ambainis}.
The walker, which is a quantum particle, moves conditioned to the
result of the coin tossing operation which is realized by a
Hadamard transform~\cite{Kempe2}. For example, the transformation
for one step of the particle from an arbitrary point $X$ is simply
\begin{equation}
\label{CG}
\begin{aligned}
&|X,R\rangle\longrightarrow\frac{1}{\sqrt{2}}(|X+1,R\rangle+|X-1,L\rangle),\\
&|X,L\rangle\longrightarrow\frac{1}{\sqrt{2}}(|X+1,R\rangle-|X-1,L\rangle).
\end{aligned}
\end{equation}
After $n$ steps, the state of the system is $\ket{\Psi_{n}}$.
Differently from the classical walks on a line, where the
position of the particle is monitored at every step of the
process, in the quantum
 version the walker remains in a superposition of many positions until the final measurement is performed.
 The probability for the particle being at $X_k$ after $n$ steps is
${\cal P}_n(X_k)=|\langle R|\langle X_k|\Psi_n\rangle|^2+|\langle L|\langle X_k|\Psi_n\rangle|^2$.
%\begin{equation}
%{\cal P}_n(X_k)=|\langle R|\langle X_k|\Psi_n\rangle|^2+|\langle L|\langle X_k|\Psi_n\rangle|^2.
%\end{equation}
During the quantum random walk process, destructive as well as
constructive interference may occur. The
quantum correlation between two different positions on a line
introduced at the first step may be kept by delaying the measurement
step until the final iteration.

The probability distribution to find
the particle at a given position is generally dependent on the
initial state of the system~\cite{Kempe2} and exhibits a very structured pattern. This allows only numerical evaluations of its variance. It has been
shown that, roughly, the standard deviation $\sigma_{QRW}$ grows
linearly with $N$ and is independent from the initial state of the
coin~\cite{Implementations}. Thus, the walker
in quantum walks explores its possible configurations faster than in
classical walks, where the standard deviation grows as $\sqrt{N}$.
This motivates the conjecture that algorithms based
on quantum random walks could beat their classical versions in terms
of the time needed to solve a problem~\cite{Deotto}.

There have been a few suggestions for
experimental implementations of quantum random
walks~\cite{Implementations}.  Recently, it has been shown that
quantum random walks can be realized using linear optical elements~\cite{Cinesi}.
In this scheme, polarization beam splitters, half-wave plates and photodetectors are used.
The walker is embodied in a single-photon
state and the entire scheme is based on the quantum coherence of two polarization states of the photon.
This result is inspiring as a first
proposal for an all-optical implementation of a quantum random
walk, even if it requires a reliable single-photon state source, which is
very demanding, and the apparatus is highly sensitive to
variations in the photons polarization.

First, we propose a scheme which uses ordinary 50:50 beam
splitters, phase shifters and photodetectors. We formulate
quantum random walks with the coin tossing operation embedded in
the translation of the walker particle. In our scheme, the
polarization degree of freedom does not play a role and, thus, is not considered at all.
A single-mode field, including a thermal field,
 may be used as an input to simulate the distribution of the quantum random walk. In
fact, this may be apparent if we recall that Young used a thermal
field for his double-slit experiment and showed interference.
\begin{figure} [ht]
\centerline{\psfig{figure=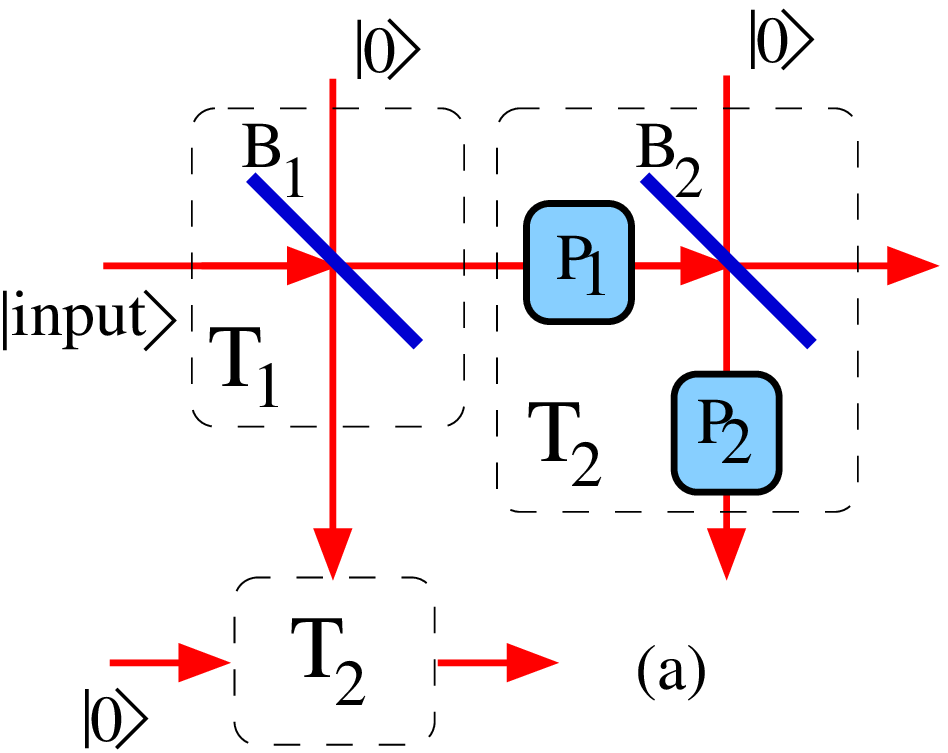,width=6.5cm,height=5.0cm}}
\vspace{0.7cm}
\centerline{\psfig{figure=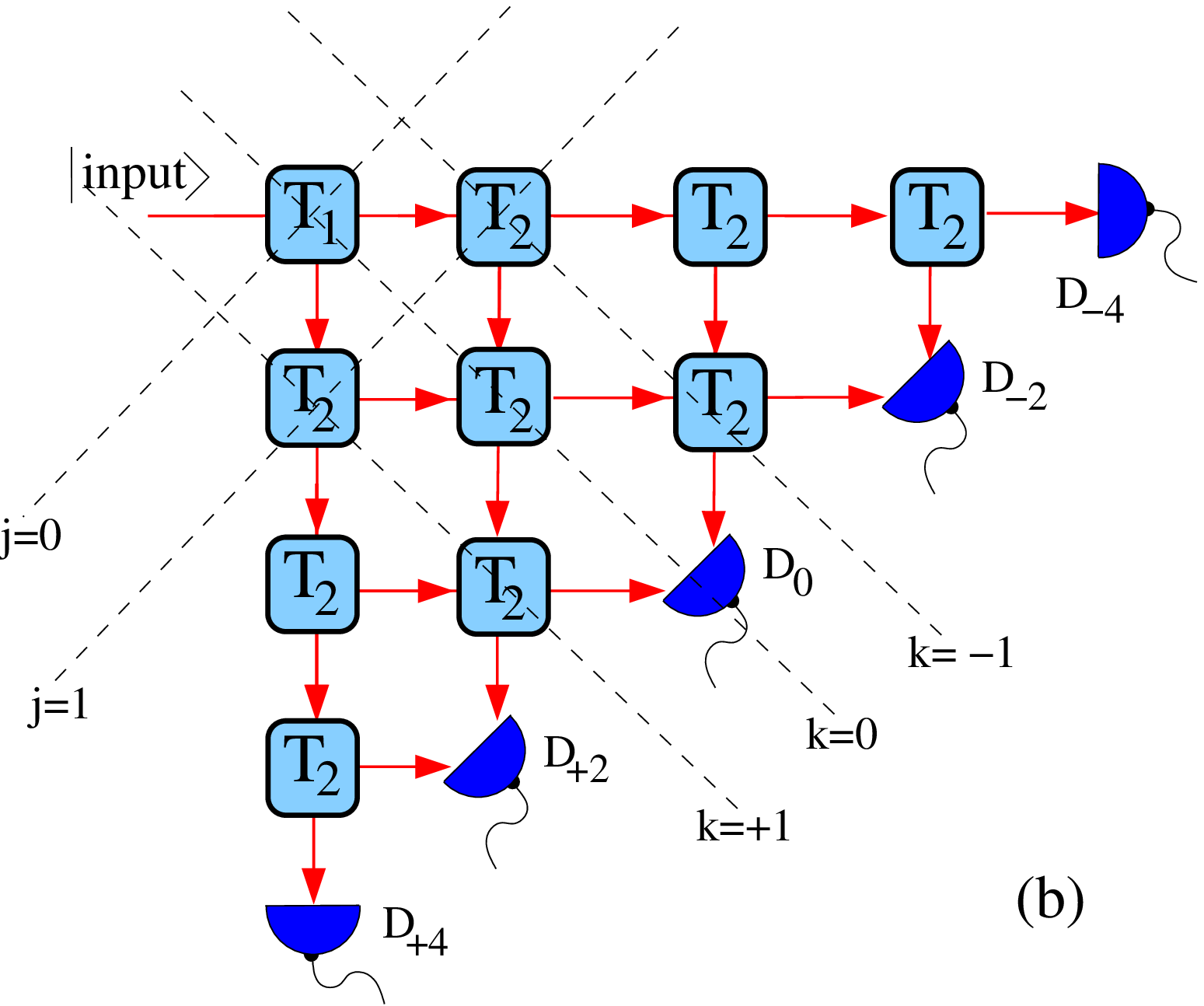,width=8.5cm,height=7.5cm}}
\caption{All-optical set-up for the simulation  of quantum
random walks on a line. ({\bf a}) Two different kinds of
  operations are shown: $\hat{T}_{1}$ is an ordinary
  beam splitter $\hat{B}_1(\theta,\phi)$.  $\hat{T}_{2}$
  involves the cascade of the phase shifter $\hat
  {P}_{1}(\pi/2)$, of a 50:50 beam splitter
  $\hat{B}_2(\pi/2,\pi)$ and of the phase shifter
  $\hat{P}_{2}(-\pi/2)$. ({\bf b}) Proposed set-up,
  shown up to the fourth dynamic line. Apart the input state, all the
  other modes are initially prepared in vacuum states.}
\label{set-up}
\end{figure}

Let us consider the experimental set-up, composed of 50:50 beam
splitters, phase shifters, and photodetectors, shown in
Fig.~\ref{set-up}. For convenience, we denote the field modes
propagating sidewards by $s$ and downwards by $d$. As the beam
splitters used here are polarization insensitive, these modes do
not refer to polarization. Here, we consider a single-photon
state $\ket{1}_{s}$ as the initial state of the walker and we
show that, in this case, our scheme gives rise to coined quantum
random walks on a line. At the first beam splitter,
$\hat{B}_1(\theta,\phi)$, the input field is mixed with a field
mode prepared in a vacuum state  (Fig. \ref{set-up}({\bf a})).
The following transformation is realized
\begin{equation}
\label{B1}
\hat{B}_1(\theta,\phi)|0,1\rangle_{ds}=
\cos\frac{\theta}{2}|1,0\rangle_{ds}
+e^{i \phi}\sin\frac{\theta}{2}|0,1\rangle_{ds}
,
\end{equation}
where $\hat{B}_1(\theta,\phi)=\exp\{\theta/2(e^{i\phi}
\hat{a}^{\dagger}_{s}\hat{a}_{d}-e^{-i\phi}
\hat{a}_{d}^{\dagger}\hat{a}_{s})\}$ is the beam splitter
operator and $\hat{a}_{s,d}$ ($\hat{a}^{\dagger}_{s,d}$) are the
annihilation (creation) operators for a sideward and a downward
field mode, respectively. We define the transformation in Eq.~(\ref{B1}) as $\hat{T}_1$. We introduce the
transformation $\hat{T}_2$
\begin{eqnarray}
&&|1,0\rangle_{d,s}\rightarrow\frac{1}{\sqrt{2}}(|1,0\rangle+|0,1\rangle)_{d,s},\\
&&|0,1\rangle_{d,s}\rightarrow\frac{1}{\sqrt{2}}(|1,0\rangle-|0,1\rangle)_{d,s},
\end{eqnarray}
which can be realized with a 50:50 beam splitter, $\hat{B}_{2}(\pi/2,\pi)$,
and two phase shifters $\hat{P}_1(\pi/2)=e^{i\pi\hat{a}^{\dagger}_{s}\hat{a}_{s}/2}$ and
  $\hat{P}_2(-\pi/2)=e^{-i\pi\hat{a}^{\dagger}_{d}\hat{a}_{d}/2}$ as shown in Fig.~\ref{set-up}({\bf a}).

The scheme can simply be illustrated as recursive applications of
$\hat{T}_2$ after the initial transformation $\hat{T}_1$, as
shown in Fig.~\ref{set-up}({\bf b}). A dynamic line \cite{Cinesi}
is represented by a row of aligned optical elements (or
photodetectors), labelled $j$ in Fig.~\ref{set-up}({\bf b}). On
the other hand, a node is given by a point represented by $k$ on
a dynamic  line. For example, the detector $D_{-2}$ is on the
fourth dynamic line and occupies the node $k=-2$. If a photon is
incident downward (sideward) on a dynamic line $j$ and node $k$,
we represent its state as $|k,d\rangle_j$ ($|k,s\rangle_j$). The
transition from a dynamic line $j$ to $j+1$ by means of the
operation $\hat{T}_2$ is synthesized by
\begin{equation}
\label{QuantumG}
\begin{aligned}
&|k,d\rangle_j\rightarrow\frac{1}{\sqrt{2}}\left(|k+1,d\rangle+|k-1,s\rangle\right)_{j+1},\\
&|k,s\rangle_j\rightarrow\frac{1}{\sqrt{2}}\left(|k+1,d\rangle-|k-1,s\rangle\right)_{j+1}.
\end{aligned}
\end{equation}
We notice that %Eq.~(\ref{QF}) is equivalent to Eq.~(\ref{CF})
Eqs.~(\ref{QuantumG}) are equivalent to Eqs.~(\ref{CG}). Thus, the
actions of $\hat{T}_1$ and $\hat{T}_2$ on a single-photon state
exactly corresponds to a coined quantum random walk. Any initial
coin state, up to an irrelevant global phase, can be prepared
changing $\theta$ and $\varphi$ in $\hat{T}_{1}$. If
$\theta=\pi/2$ and $\phi=-\pi/2$, we get the symmetric
probability distribution that corresponds to the initial coin
state $(|R\rangle+i|L\rangle)/\sqrt{2}$ in a coined quantum
walk~\cite{Implementations}. In our model, the difference between
quantum and classical walks from a certain step is due to the
interference of the walker's paths on the $\hat{T}_2$ processes~
\cite{footnote}.

%%%%%%%%%%%%%%%%%%%%%%%%%%%%%%%%%%%%%%%%%%COHERENT STATES%%%%%%%%%%%%%%%%%%%%%%%%%%%%%%%%%%%%%%%%%%%%%%%%%%%%%%%%%%%%%%%
\section{Analysis with different states of the walker}
\label{general proof}
In this Section we show that the scheme suggested in Fig.~\ref{set-up} exhibits the same
interference pattern at the detectors regardless of the nature of the input state. We first address the case of an input
coherent state and, then, we extend the analysis to any field.
%%%%%%%%%%%%%%%%%%%%%%%%%%%%%%%%%%%%%%%%%%%%%%%%%%%%%
\subsection{Coherent states}
\label{coherentstates} 
\begin{figure}
\centerline{\psfig{figure=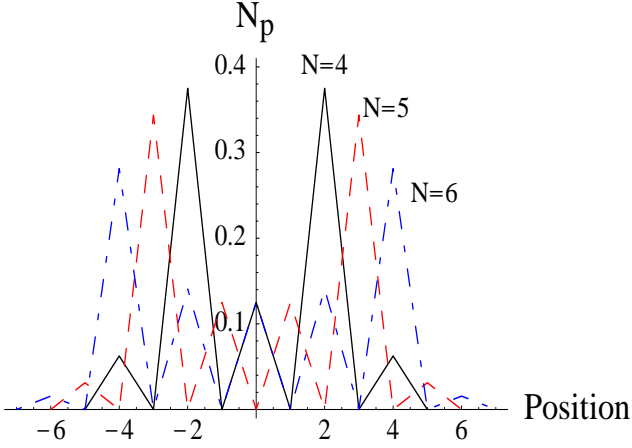,width=8.4cm,height=6.0cm}}
\caption{Average photon-number distribution  for an input
coherent state $\ket{\alpha=1}$, as a function of the position
along the final dynamic line. Three different cases are
considered: the solid-line curve is relative to a number of steps
$N=4$; the dashed-line represents $N=5$ while the dot-dashed one
is for $N=6$. The plots match perfectly the graphs expected for a
coined quantum walk on a line. In the general case of
$\alpha\neq{1}$, $N_{p}$ has to be normalized with respect to
$\modul{\alpha}^2$.}
\label{coherentstate}
\end{figure}

A coherent state $\ket{\alpha}$
($\alpha\in\mathbb C$) is generally assumed to be the best
description of the state of a laser beam. We consider
$\ket{\alpha}$ as the input state of the walker. The action of
the beam splitter operator on two input coherent states does not
lead to any entanglement between the output modes~\cite{kim}.
Assuming $\theta=\pi/2$ and $\phi=-\pi/2$ for the $\hat{T}_1$
process, we can calculate the distribution of the average
photon-number as a function of the position $k$ on the chosen
final dynamic line. For example, for $N=4$ steps, we find the
final state
\begin{equation}
\label{quintostep}
\begin{aligned}
\ket{\Phi_{4}}&=\ket{\frac{-i\alpha}{4},s}^{-4}_{4}
\ket{\frac{-i\alpha}{4},d}^{-2}_{4}\ket{\frac{1-2i}{4}\alpha,s}^{-2}_{4}\ket{\frac{\alpha}{4},d}^{0}_{4}\\
&\otimes\ket{\frac{i\alpha}{4},s}^{0}_{4}\ket{\frac{-2-i}{4}\alpha,d}^{+2}_{4}
\ket{\frac{\alpha}{4},s}^{+2}_{4}\ket{\frac{-\alpha}{4},d}^{+4}_{4},
\end{aligned}
\end{equation}
with $\ket{\alpha,s}^{k}_{j}$ ($\ket{\alpha,d}^{k}_{j}$) that indicates a coherent state incident sideward
(downward) on a dynamic line $j$ and node $k$. The average photon-number ${\cal N}_{p}(N,k)$ for node $k$ is
${\cal N}_{p}(4,k)={\cal M}(4,k){\cal N}_{in}(|\alpha\rangle)$, with
\begin{equation}
{\cal M}(4,\pm{4})=\frac{1}{16},~ {\cal
M}(4,\pm{2})=\frac{3}{8},~ {\cal M}(4,0)=\frac{1}{8}.
\end{equation}
Here, ${\cal N}_{in}(|\alpha\rangle)=\modul{\alpha}^2$ is the
average photon number for the input state $|\alpha\rangle$ and
${\cal M}(N,k)$ is the {\it normalized photon-number
distribution} at step $N$ and node $k$. It characterizes the
output photon-number distribution at the detectors. We find that
the distribution ${\cal M}(4,k)$ for an input coherent state is
the same as the one for the single photon
input~\cite{Implementations}, {\it i.e.}, the two different
inputs result in the same photon-number distribution. The average
photon numbers for steps, $N=4,5,6$ are shown in
Fig.~\ref{coherentstate}. The deviations of a quantum walk from
its classical counterpart appears from the fourth step. This is
due to the particular values of the parameters in the
transformation $\hat{T}_{1}$: $\theta=\pi/2$ and $\phi=-\pi/2$.
Since a coherent state input results in the same quantum random
walk pattern of the single photon case for all the steps we have
considered, we conjecture that the quantum walk pattern results
even any initial state for a general number of steps $N$. In what
follows, we prove the validity of this conjecture.

\subsection{General case}
\label{Proof}
With the proposed set-up, the quantum walk process can be
represented as
\begin{eqnarray}
|\Phi_N\rangle={\hat{U}}_{T{(j=N)}}...{\hat{U}}_{T(j=1)}{\hat{T}}_{1(j=0)}
|\Phi_0\rangle\nonumber\equiv&\widehat{U}^{N}_{QW}|\Phi_0\rangle,
\end{eqnarray}
where $\ket{\Phi_{0}}$ is the input state, $N$ is the number of
steps, and $\hat{U}_T$ is an appropriate unitary transformation for
each step. For a coherent state, the previous result can be
summarized as
\begin{equation}
\label{evolution}
|\Phi_0\rangle=
|\alpha\rangle
\stackrel{{\widehat U}^{N}_{QW}}\longrightarrow
|\chi_1\alpha\rangle_1|\chi_2\alpha\rangle_2...
|\chi_N\alpha\rangle_{2N}=|\Phi_N\rangle.
\end{equation}
Eq.~(\ref{quintostep}) is an explicit example. The average photon
number for mode $r$ ($0\le{r}\le{2N}$) is
$n_{r}=|\chi_r|^2|\alpha|^2 =|\chi_r|^2{\cal
N}_{in}(|\alpha\rangle)$,  with $r=0$ corresponding to the mode
incident on the detector that occupies $j=N$, $k=-N$. It is easy
to show that the average photon number for the $k$-th node and
$j$-th step is given by ${\cal N}_p(j,k)
=n_{j-k}+n_{j-k+1}=(|\chi_{j-k}|^2+|\chi_{j-k+1}|^2){\cal
N}_{tot}(|\alpha\rangle)$, where $\chi_0=\chi_{2N+1}\equiv0$.
This result also means that
\begin{equation}
\label{also}
{\cal M}(j,k)=|\chi_{j-k}|^2+|\chi_{j-k+1}|^2.
\end{equation}
 Note that $\chi_r$ does not depend on the amplitude of
the initial state but only on the structure of $\widehat{U}^{N}_{QW}$.

The initial state density operator in $P$ representation can be
generally written as \cite{MandelWolf,ScullyZubairy}
\begin{equation}
\label{p}
\rho_0=\int d^2\alpha P(\alpha)|\alpha\rangle\langle\alpha|,
\end{equation}
where $P(\alpha)$ is the $P$ representation of the initial state $\rho_0$.
Provided that $P(\alpha)$ is a sufficiently singular generalized
function, such a representation exists for any given operator
$\rho_{0}$~\cite{ScullyZubairy}. After $N$ steps, the density operator
evolves as:
\begin{equation}
\label{prova}
\begin{aligned}
\rho_N&=\hat{U}^{N}_{QW}\rho_{0}\hat{U}^{N^{\dagger}}_{QW}\\
&=\int d^2\alpha P(\alpha)|\chi_1\alpha\rangle_1
\langle\chi_1\alpha|\otimes..\otimes|\chi_{2N}\alpha\rangle_{2N}\langle\chi_{2N}\alpha|
\end{aligned}
\end{equation}
where Eqs.~(\ref{evolution}) and (\ref{p}) have been used. The $P$ representation is
particularly appropriate for our aim to find the average photon-number
distribution since it can be shown that the moments of the $P$ representation
give the expectation values of normally-ordered products of
bosonic operators~\cite{MandelWolf,ScullyZubairy}.

The marginal density matrix for mode $r$ is simply obtained as
\begin{equation}
\begin{aligned}
\rho_r%{\rm Tr}_{1,2...r-1,r+1,r+2...2N-1,2N}\left(\rho_N\right)\nonumber\\
=\int d^2\alpha
P(\alpha)|\chi_r\alpha\rangle_r\langle\chi_r\alpha|.
\end{aligned}
\end{equation}
The average photon number for the $r$-th mode is
\begin{equation}
\nonumber
n_{r}={\rm Tr}_{r}[\rho_r \hat{a}^{\dagger}\hat{a}]=|\chi_r|^2\int d^2\alpha
P(\alpha)\modul{\alpha}^2=|\chi_r|^2{\cal N}_{in}(\rho_0),
%\end{aligned}
\end{equation}
and the average photon number for the $j$-th step and $k$-th node
is ${\cal N}_p(j,k) ={\cal M}(j,k){\cal
N}_{tot}(\rho_0)=(|\chi_{j-k}|^2+|\chi_{j-k+1}|^2){\cal
N}_{tot}(\rho_0)$, from which Eq.~(\ref{also}) is found to hold
for the case of any input field. The interference pattern
determined by ${\cal M}(j,k)$ does not depend on the initial
input state. For a given set of beam splitters and phase
shifters, any input state will result in the same interference
pattern. Only an overall factor will be changed, according to the
total average photon-number of the initial state. For a classical
light, in a pictorial way, the result is nothing but quantum
random walks with many walkers simulated by interference between
fields. For a weak field, the quantum random walks with a single
walker can be probabilistically performed. For example, given a
coherent state with $\alpha=1$, a single photon is detected with
$37\%$ of the probability.

A problem of the approach employing {\it dynamic lines} for
quantum random walks is that the required number of resources
(in terms of the number of optical elements required for a chosen
number of steps and of the field modes involved) grows
quadratically with the number of steps. This imposes serious
limitations to the scalability of such a proposal and affects the
efficiency of a simulation based on an interferometric device.
\begin{figure}[ht]
\centerline{\psfig{figure=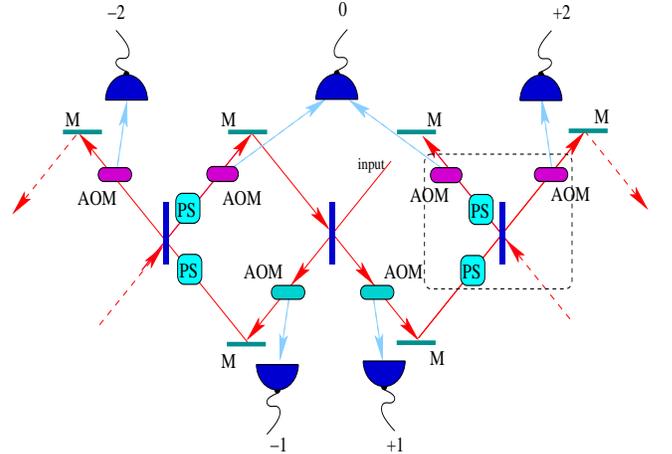,width=8.5cm,height=6.0cm}}
\caption{Alternative set-up for quantum random walk on a
line. In this scheme, the number of required resources scales linearly with the nymber of steps $N$.
Two rows of Acousto-Optic Modulators (AOM) direct the incoming
beams of light to the perfect mirrors M or to the detectors row.
This set-up is conceptually equivalent to that sketched in
Fig.~\ref{set-up}({\bf b}).}
\label{set-up2}
\end{figure}
In the alternative proposal in Fig.~\ref{set-up2}, this problem
is bypassed measuring all the even positions by the upper row of
detectors, while the odd ones are detected by the lower row.
Acousto-Optic Modulators (AOMs) \cite{aom} are used to guide a
beam toward a mirror for further steps or toward a detector for
the measurement. When the AOMs in the top row have to deflect the
light beams toward the detectors, those in the bottom row should
not be active. The beam splitters and phase shifters in
Fig.~\ref{set-up} and in Fig.~\ref{set-up2} are the same. The
number of required resources, in this latter scheme, increases
only linearly with the number of steps~\footnote{More precisely,
while the number of beam splitters grows as $N$, the required
AOMs and phase shifters increase as $2N$. However, in the spirit
of the proposal shown in Fig.~\ref{set-up}, we can consider basic
building blocks made by a beam splitter, two phase shifters and
two AOMs (as the one boxed in Fig.~\ref{set-up2}). This clarifies
that, in this case, the number of building blocks grows linearly
with $N$.}.

There are many difficulties,  for a practical implementation of
the schemes we propose, that have to be taken in consideration.
For example, being a multimode interferometric apparatus, the
proposed set-up could be affected by the misalignment of the
involved optical elements. Furthermore, we need $2N$ modes for
$N$ steps of the walk process, that makes the controllability of
the system very difficult. Nevertheless, even if these problems
render the proposed set-up challenging under an experimental
point of view, our proposal has to be seen as a thought
experiment useful for the investigation of the physics that is
behind the appearance of the characteristic probability pattern
of a quantum random walk.

%%%%%%%%%%%%%%%%%%%%%%%%%%%%%%%%%%%%%%%%%%%%%%%%%DECOERENZA%%%%%%%%%%%%%%%%%%%%%%%%%%%%%%%%%%%%%%%%%
\section{Simulation of decoherence in quantum random walks}
\label{decoerenza}

To better understand how interference effects are at the basis of
a quantum walk process we study the effect that a certain class
of errors have on the performance of the set-up we propose. A
decoherence mechanism is potentially able to wash out the
interference pattern, thus  erasing the speed-up characteristic
of a quantum walk and restoring some aspects of the classical
diffusion process. In this Section we study two different models
for decoherence in our set-up. We show the transition of the
dynamics of the walker from the pure quantum to the classical
case. This analysis is in part motivated by the attention that
has been recently payed to the way in which the quantum walk
pattern is modified by imperfect coin tossings or walker
translations, both for quantum walk on a line and higher
dimensions~\cite{Implementations,decokendon}. A remarkable
result, shown by Kendon and Tregenna in \cite{decokendon}, is
that small amounts of decoherence, rather than render the process
useless for the purposes of quantum information, amazingly
increase the capability of the system to explore its possible
configurations. This gives a probability distribution to find the
walker in a certain position that spreads faster than in pure
dynamics. Our study is able to highlight even this aspect of the
dynamics of the walker. On the other hand, studying the effects
of possible sources of decoherence is worth under a practical
point of view. The characterization of some relevant sources of
errors, in the proposed set-up, will make us understand why the
pattern resulting from a performed experiment devoted to the
realization of a quantum walk process could deviate from the
ideality.

We have considered ulterior phase shift operations performed just
before and after each $\hat{T}_{2}$ transformation. The shift in
these additional operations is randomly chosen from a Gaussian
distribution. In what follows, we show how the mean photon-number
distribution changes its shape (from a classical Gaussian
pattern to an approximately flat distribution then to a quantum distribution)
as the amount of randomness in the additional phase shifts is reduced.

If $l$ is a number randomly taken from a Gaussian distribution
centred at $1$ with an adjustable standard deviation
$\sigma_{pp}$, we shift the phase of each field mode in
Fig.~\ref{set-up}({\bf b}) by an amount equal to $2\pi\modul{l}$.
If the phase shift is equal to $2\pi$, the additional phase
shifters are ineffective and a quantum walk pattern is recovered.
On the other hand, if the amount of shifts deviates from this
{\it neutral} value, they affect the interferences responsible
for the quantum walk and some deviations have to be expected and
the average over a  large number of trials results in a classical
distribution.
\begin{figure} [ht]
\centerline{\psfig{figure=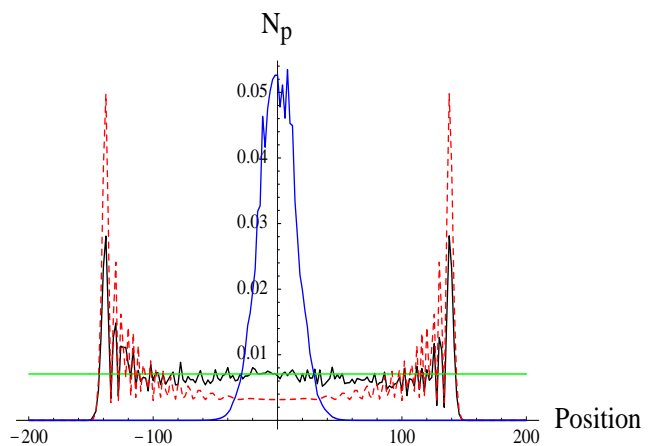,width=8.5cm,height=6.0cm}}
\caption{Average photon-number distribution vs position for an
input coherent state $\ket{\alpha=1}$ and for 200 steps.
Different cases are considered: the bell-like curve represents
the case of an introduced randomness $l$ taken from a Gaussian
distribution centred at $1$ and with $\sigma_{pp}=0.25$. The
curve evidently resembles the expected Gaussian distribution. The
solid line shows the results for $\sigma_{pp}=0.0125$. It is compared with a uniform
distribution
between $-200/\sqrt{2}$ and $200/\sqrt{2}$, fictitiously extended to improve visibility.
Finally, the dashed curve represents the pure quantum case
corresponding to $l$ chosen from a Dirac delta function
$\delta(l-1)$. Each point in the simulated
curves is averaged over 50 different trials.}
\label{deco}
\end{figure}

In Fig.~\ref{deco}, the shown distributions are the results of an
average over $50$ different trials: in each one of them, and for
each step in a single trial, a different random value for $l$ is
considered and the mean photon-number at the various locations on
the final dynamic line is calculated, averaging over the outcome
for each trial.

If, now, $\sigma_{pp}$ is reduced (in Fig.~\ref{deco},
$\sigma_{pp}=0.0125$), the phase shifts vary over a small range
of values around $2\pi$. The dynamic evolution of the system is
affected in such a way that no classical signature is evident in
the mean photon-number distribution. A highly non-classical
pattern is found and some deviations from the pure quantum random
walk case are evident. The distribution is relatively flat over a
region that is wider than the pure quantum case. This result is
in good agreement with the analysis performed
in~\cite{decokendon} for a small amount of decoherence. In our
case, the limited randomness imposed to the evolution of the
photonic walker simulates the effect of a decoherent coin tossing.
%As long as $l$ stays near to 1, the
%walk process not only keeps the characteristics of a quantum
%distribution but even spreads faster.
The remarkable feature in this analysis is that we have used just
classical resources (linear optics elements and input coherent
states). Nonetheless, we still simulate the relevant features of the
transition from a pure quantum evolution to the
classical spread due to a large superimposed randomness.

\begin{figure} [ht]
\centerline{\psfig{figure=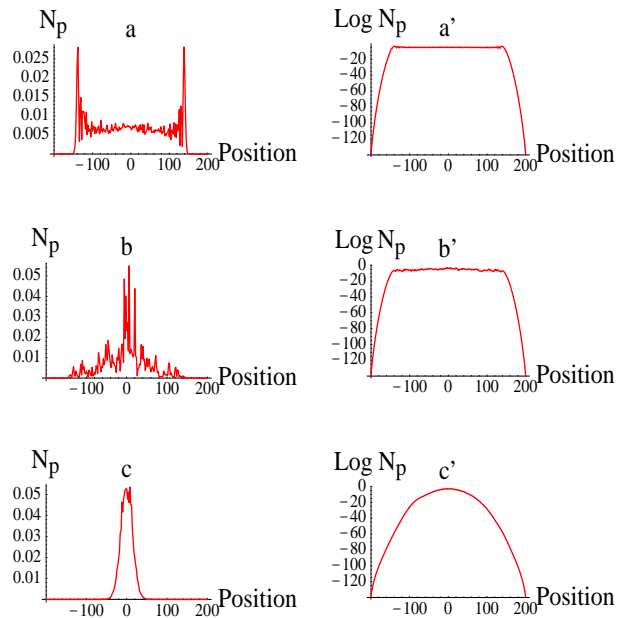,width=8.5cm,height=9.0cm}}
\caption{Transition from weak to strong randomization in the
model for decoherence in the coin tossing for an input coherent
state $\ket{\alpha=1}$. From top to bottom, $\sigma_{pp}$, is
increased. We have considered $\sigma_{pp}=0.013$ ({\bf a}),
$\sigma_{pp}=0.13$ ({\bf b}) and $\sigma_{pp}=0.25$ ({\bf c}).
The figures in the right show the same distributions presented in
the left but in lin-log scale, with which the investigation of the appearance of
localization effects is easier. The mean photon-number distribution smoothly changes from
a sharp-squared distribution to a concave curve that is typical
of a classical distribution~\cite{decojex}.}
\label{Anderson}
\end{figure}

In~\cite{decojex}, the effect of phase randomness in a general
interferometric device has been investigated. In particular, if
the device can be thought as the iterative applications of some
basic units, each one affected by a fixed
randomness~\cite{decojex}, then Anderson localization can be
obtained. Indeed, when fixed randomness is considered, a
connection to the theory of the band-diagonal transfer matrix
(examined in~\cite{haake}) can be established. It is this kind of
dynamic evolution that leads to localization of the walker.
Physically, the model described in this case is near to the
repeated passages of a beam of light through a dielectric layer
placed inside an electro-magnetic cavity, as described
in~\cite{dik}. In our model, however, no localization effect is
achieved since different values for the phase shifts at each step
are taken. In this respect, our case is far from a band-diagonal
evolution. These qualitative arguments are resumed in
Fig.~\ref{Anderson}, where the transition from a flat
distribution (obtained for a small decoherence parameter
$\sigma_{pp}$) to the classical one (relative to a strongly
randomized quantum walk) is reported. To compare our results to
those in~\cite{decojex} and to show that no dynamic localization
is here achieved, we present plots for the average photon-number
distributions in lin-log and in lin-lin scale.

Following the same lines depicted above, we can investigate about
errors due to the uncertainty in the beam splitters
transmittivities. We consider imperfect beam splitters whose
transmittivities randomly fluctuate around $50\%$ according to a
Gaussian distribution with standard deviation $\sigma_{bs}$.
Computing the normalized average photon-number distribution for
an input coherent state, we find a narrow range of values for
$\sigma_{bs}$ within which a flat distribution is achieved.
Outside this range, the distribution rapidly converges toward a
classical one.

\begin{figure} [ht]
\centerline{\psfig{figure=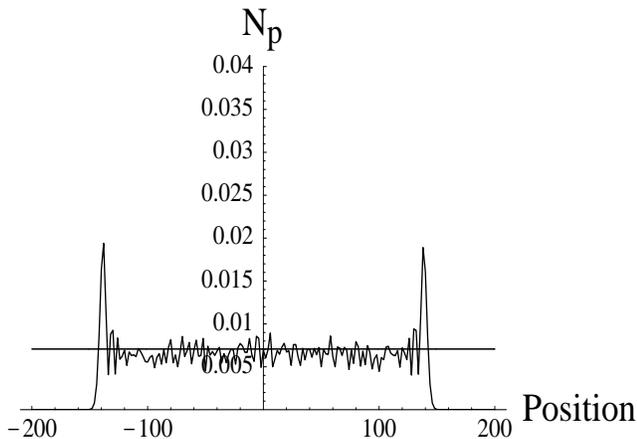,width=8.5cm,height=6.0cm}}
\caption{Average photon-number distribution for an input state
$\ket{\alpha=1}$ considering both the models of
decoherence. The number of steps considered
is $200$ and each point is averaged over 50
different trials. We have taken $\sigma_{pp}=0.005$. On the other
hand, we have taken $\theta=\frac{\pi}{2}\modul{m}$ in $\hat{T}_{2}$, with a random
number, $m$, extracted from a Gaussian distribution centered at $1$ and
having standard deviation $\sigma_{bs}=0.07$.}
\label{totale}
\end{figure}

To give a picture of the combined effect of the two decoherence
processes, we include random phase shifters between two
subsequent $\hat{T}_{2}$ operations and random fluctuations in the
transmittivity of the beam splitters. In Fig.~\ref{totale} we
show the distribution that corresponds to $\sigma_{pp}=0.005$ and
$\sigma_{bs}=0.07$. We can see that the mean-photon  distribution
has been very much flattened. Of course, as $\sigma_{pp}$ and
$\sigma_{bs}$ grow, the curve will become Gaussian.

%%%%%%%%%%%%%%%%%%%%%%%%%%%%%%%%%%REMARKS%%%%%%%%%%%%%%%%%%%%%%%%%%%%%%%%%%%%
\section{Remarks and Discussion}

As we discussed, the realization of the model we propose is  not
trivial as we pays the price represented by the use of $2N$ field
modes to replace the quantum walker (that belongs to a Hilbert
space of dimension $N$). Thus, the addition of a component, in
our set-up, increases the difficulty of alignment. However, what
we want to stress in this paper is the possibility of simulating
quantum random walks using the wave nature of a classical field.
We have shown that this study has been possible using our thought
experimental set-up. Even though this results could be
surprising, the possibility of such a simulation may be a natural
result if we consider that quantum coherence and quantum
interference are concept originally borrowed from wave mechanics.
This possibility has been formally proved using standard tools of
quantum optics.

Furthermore, we have simulated some decoherence mechanisms on the
quantum random walk by means of linear optical devices and input
coherent states. We have observed how the average photon-number
distributions are modified when controlled randomness is
introduced in the system via additional phase shifters and
imperfect beam  splitters. This analysis is useful both
theoretically (clarifiying the role of the coherent effects in
the simulation) and practically because it characterizes the
influences of possible sources of errors affecting the results of
a performed experiment.

\begin{figure} [ht]
\vspace{0.2cm}
\centerline{\psfig{figure=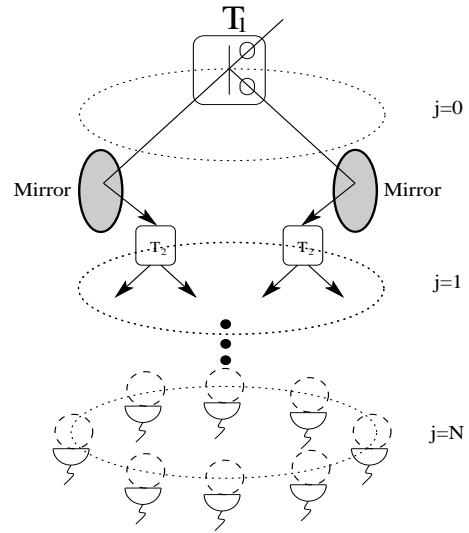,width=6.0cm,height=7.0cm}}
\caption{An implementation of a quantum random walk on a circle
using
  \it{dynamic circles}.}
\label{circle}
\end{figure}

Finally, we want to mention here that it is in principle possible
to extend our scheme to quantum random walks on a circle of $N$
points, as shown schematically in Fig.~\ref{circle}.  One can
adapt the concept of dynamic line to that of {\it{dynamic
circles}}: the walker transits from circle to circle (each having
a non decreasing number of sites on it) in a way completely
similar to that described in Section~\ref{Setup}. The number of
required dynamic circles is equal to $N$. Each site on a given
circle is occupied by a basic operation: $\hat{T}_{1}$ occupies
the unique site on the first dynamic circle, all the other sites
in the following circles (labeled as $j=1,..,N$ in
Fig.~\ref{circle}) being occupied by $\hat{T}_{2}$. After each
$\hat{T}_{2}$ operation, the beams are directed, by means of some
mirrors, toward the proper site on the next dynamic circle, as
shown in Fig.~\ref{circle} for the transition from the $j=0$ to
the $j=1$ circle. At the final dynamic circle, the mean-photon
number distribution at the sites is revealed by an array of
detectors. Basically, this implementation is still based on the
simulation of a quantum walk on a line and it is, thus, obvious
that it will simulate quantum walks on a circle with classical
fields.

Under certain circumstances, our approach can be useful in order
to simulate quantum walks on a hypercube of dimension 3. This
higher-dimensional quantum walk can, indeed, be reduced to a
biased quantum walk on a line with properly chosen, asymmetrical,
probabilities for the coin to be in the $\ket{R}$ or in the
$\ket{L}$ state~\cite{Kempe2}. As we have seen in
Section~\ref{Setup}, properly choosing the parameters of the
optical elements in $\hat{T}_{1}$, $\hat{T}_{2}$, our proposal is
able to realize quantum random walk on a line with any biased
coin. We, thus, expect the possibility to simulate quantum walks
on a three-dimensional hypercube by means of interference of
classical light. However, the extension of these results to
general graphs and hypercubes of higher dimensions (as well as an
analysis of the efficiency of such a simulation) is much more
difficult and goes beyond the purposes of this work. It is,
however, worth stressing that the possibility of classical
simulations of quantum random walk on a line does not necessarily
imply their usefulness for a practical quantum algorithm. There
remain important open questions about the gain, in terms of
speed-up of quantum computation, that can be obtained from the
classical simulation of quantum walks.

%%%%%%%%%%%%%%%%%%%%%%%%%%%%%RINGRAZIAMENTI E FINANZIAMENTO%%%%%%%%%%%%%%%%%%%
{\bf Note added -} Knight {\it et al.} also pointed out the
possibility of simulation of quantum random walks using classical
fields but using a totally different set-up~\cite{Knight03}. This appeared one day before the present
manuscript of ours was uploaded into the Los Alamos archive.

\acknowledgments

We thank V. Kendon, G. M. Palma, I. A. Walmsley and Z. Zhao for stimulating discussions and
useful comments. This work has been
supported by the UK Engineering and
Physical Science
Research Council grant GR/S14023/01. M.P. acknowledges IRCEP for financial
support.


\begin{thebibliography}{99}

\bibitem{MandelWolf} L. Mandel and E. Wolf, {\sl Optical coherence and quantum optics} (Cambridge University  Press,
1995).

\bibitem{Vigoda} U. Sch\"oning, {\sl  Proceedings of the {40th} Annual
Symposium on Foundations of Computer Science}, New York, NY, 1999;
M. Jerrum, A. Sinclair, E. Vigoda, {\sl Proceedings of the {33rd}
ACM Symposium on Theory of Computing}, 2001.

\bibitem{EF} R. A. Dana and M. Jeanblanc, {\sl Financial Markets in continuous time}, (Springer, Berlin, 2002).

\bibitem{Ambainis} A. Ambainis, E. Bach, A. Nayak, A. Vishwanath,
J. Watrous, {\sl Proceedings of 33rd STOC} (Assoc. for Comp. Machinery, New York, 2001);
D. Aharanov, A. Ambainis, J. Kempe, U. Vazirani, {\sl ibidem}; J. Kempe, quant-ph/0205083.

\bibitem{Kempe2} J. Kempe, quant-ph/0303081 (to appear in {\sl Contemporary Physics}).

\bibitem{Implementations} B. C. Travaglione, G. J. Milburn, {\sl Phys. Rev. A} {\bf
    65}, 032310 (2002); W. D\"ur, R. Raussendorf, V. M. Kendon,
    and H.-J. Briegel, {\sl Phys. Rev. A} {\bf 66}, 052319 (2002);
    B. C. Sanders, S. D. Bartlett, B. Tregenna, and P. L. Knight, {\sl Phys. Rev. A} {\bf 67}, 042305 (2003).

\bibitem{Cinesi} Z. Zhao, J. Du, H. Li, T. Yang, Z.-B. Chen, and J.-W. Pan, quant-ph/0212149.

\bibitem{bergou} M. Hillery, J. Bergou, and E. Feldman, quant-ph/0302161.

\bibitem{Deotto} A. M. Childs, E. Deotto, E. Farhi, J. Goldstone,
S. Gutmann, and A. J. Landahl, {\sl Phys. Rev. A} {\bf 66}, 032314 (2002);
N. Shenvi, J. Kempe, K. B. Whaley quant-ph/0210064 (to appear in {\sl Phys. Rev. A}).

\bibitem{decokendon} V. Kendon, and B. Tregenna, quant-ph/0209005
(to appear in {\sl Phys. Rev. A});  T. A. Brun, H. A. Cateret,
and A. Ambainis, {\sl Phys,. Rev. A} {\bf 67}, 032304 (2003).

\bibitem{footnote} Note that classical random walks can be
easily obtained removing all the phase shifters. In this case, indeed, there can be no destructive
interference that makes the quantum random walk different from its
classical counterpart.

\bibitem{aom} A. Stefanov, H. Zbinden, N. Gisin, and A. Suarez, {\sl Phys. Rev. Lett.} {\bf88}, 120404 (2002).

\bibitem{kim} M. S. Kim, W. Son, V. Buzek, P. L. Knight, {\sl Phys. Rev. A} {\bf 65}, 032323 (2002).

\bibitem{ScullyZubairy} M. O. Scully and M. S. Zubairy, {\sl Quantum optics} (Cambridge University  Press, 1997).

%\bibitem{Gardiner} C. W. Gardiner, {\sl Quantum Noise} (Springer, Berlin, 1992).

\bibitem{decojex} P. T\"orm\"a, I. Jex, and W. P. Schleich, {\sl Phys. Rev. A} {\bf 65}, 052110 (2002).

\bibitem{haake} F. Haake, {\sl Quantum Signature of Chaos} (Springer-Verlag, Berlin, 1992).

\bibitem{dik} D. Bouwmeester, I. Marzoli, G. Karman, W. P. Schleich, and J. P. Woerdman, {\sl Phys. Rev. A} {\bf 61}, 013410 (2000).

\bibitem{Knight03} P. L. Knight, E. Roldan, and J. E. Sipe,
preprint quant-ph/0304201.

\end{thebibliography}
\end{document}